\documentclass[letterpaper,twocolumn,10pt]{article}
\usepackage{usenix,epsfig,endnotes,listings,authblk}
\usepackage{algorithm,algorithmic,amsmath}
\usepackage{tabularx,booktabs,makecell,adjustbox,multirow} 

\begin{document}

\date{}

\title{\Large \bf Crimson: Empowering Strategic Reasoning in Cybersecurity through Large Language Models}

\author[1]{Jiandong Jin}
\author[1]{Bowen Tang}
\author[1]{Mingxuan Ma}
\author[1]{Xiao Liu}
\author[2]{Yunfei Wang}
\author[1]{Qingnan Lai}
\author[1]{Jia Yang}
\author[1]{Changling Zhou\thanks{Corresponding author: zclfly@pku.edu.cn}}

\affil[1]{Peking University, Beijing, China}
\affil[2]{National University of Defense Technology, Changsha, China}

\maketitle

\thispagestyle{empty}

\subsection*{Abstract}
We introduces Crimson, a system that enhances the strategic reasoning capabilities of Large Language Models (LLMs) within the realm of cybersecurity. By correlating CVEs with MITRE ATT\&CK techniques, Crimson advances threat anticipation and strategic defense efforts. Our approach includes defining and evaluating cybersecurity strategic tasks, alongside implementing a comprehensive human-in-the-loop data-synthetic workflow to develop the CVE-to-ATT\&CK Mapping (CVEM) dataset. We further enhance LLMs' reasoning abilities through a novel Retrieval-Aware Training (RAT) process and its refined iteration, RAT-R.

Our findings demonstrate that an LLM fine-tuned with our techniques, possessing 7 billion parameters, approaches the performance level of GPT-4, showing markedly lower rates of hallucination and errors, and surpassing other models in strategic reasoning tasks. Moreover, domain-specific fine-tuning of embedding models significantly improves performance within cybersecurity contexts, underscoring the efficacy of our methodology. By leveraging Crimson to convert raw vulnerability data into structured and actionable insights, we bolster proactive cybersecurity defenses.

\section{Introduction}\label{sec:Introduction}

The cybersecurity landscape is constantly evolving, necessitating advanced strategies to understand and mitigate vulnerabilities effectively. A critical aspect of this endeavor is the integration of CVEs and Cyber Threat Intelligences (CTIs) with the MITRE ATT\&CK framework to enhance strategic reasoning and threat prediction. This paper focuses on leveraging the reasoning capabilities of LLMs to correlate CVE data with ATT\&CK tactics, techniques, and procedures (TTPs), aiming to provide a more structured and comprehensive approach to cybersecurity defense mechanisms.

\paragraph{Integrating CVE/CTI with ATT\&CK.} The core challenge in cybersecurity is not just the identification of threats and vulnerabilities through CTIs and CVEs but their effective integration with structured cybersecurity frameworks like ATT\&CK. This integration is pivotal for developing a nuanced understanding of attack vectors and enhancing the strategic reasoning capabilities of defense mechanisms. However, the process is complicated by the unstructured nature of CTI and the non-standardization of CVE descriptions \cite{poddubnyi2020vulnerability, shahid2021cvss, alam2023looking}. Recent advancements in NLP and LLMs offer promising solutions by automating the classification and interpretation of these data sources, yet the deployment of such models faces significant challenges\cite{li2021attackg, gao2022threatkg, zuber2023vox}.

The TRAM project\footnote{\label{fn:tram-url}\url{GitHub - center-for-threat-informed-defense/tram: TRAM is an open-source platform designed to advance}} , while not directly producing structured CVE mappings, exemplifies efforts to automate the mapping of CTI to ATT\&CK techniques through NER-like processes. TRAM aims to streamline the annotation of CTI reports with ATT\&CK techniques, providing foundational resources for machine learning applications in this domain. However, it is primarily designed for research purposes and highlights the complexity and potential errors in automated mapping processes.

\paragraph{Advancements in LLM Reasoning.} The emerging trend in LLMs, capable of sophisticated multi-step reasoning, presents a promising avenue for enhancing strategic reasoning in cybersecurity. Particularly, structured reasoning facilitated by LLMs can significantly improve both the interpretability and automation of threat management. When trained on cybersecurity specifics, these models can effectively analyze vast amounts of data, drawing connections between CVEs and potential ATT\&CK techniques. Yet, training these models to navigate the complex and ethical dimensions of cybersecurity requires ongoing research and development \cite{wei2022chain, kojima2022large, fu2023complexity}.

Our research contributes to this field by proposing a novel framework that not only maps CVEs to ATT\&CK techniques but also enhances the strategic reasoning capabilities of LLMs within the cybersecurity context. By doing so, we aim to bridge the gap between technical vulnerability details and practical cyber threat aspects, enabling a proactive approach to cybersecurity.

\paragraph{Contributions.} Our main contributions include:
\begin{itemize}
    \item We designed a task for assessing strategic reasoning in cybersecurity and created an comprehensive dataset through a synthetic data workflow. To our knowledge, this is the first dataset aimed at tasks related to cybersecurity strategic reasoning capabilities.
    \item We implemented Retrieval-Aware Training (RAT) and its enhanced variant, RAT-R, which boost the strategic reasoning capabilities of smaller models, enhancing their performance in generating precise strategies.
    \item We conducted an extensive evaluation, proving that our model, closely rivals the capabilities of leading LLMs like GPT-4 while demonstrating a reduced propensity for hallucinations and errors. Significantly surpasses other open-source models in strategic reasoning ability.
    \item We highlighted the critical role of domain-specific fine-tuning of embedding models in improving reasoning accuracy within cybersecurity contexts, thereby proving the effectiveness of our methodologies.
\end{itemize}

The remainder of this paper is organized as follows: Section~\ref{sec:Background} provides an overview of CVEs, CTIs, and the MITRE ATT\&CK framework, setting the groundwork for our study. Section~\ref{sec:Methodology} describes our methodology, including dataset development and model fine-tuning processes. Section~\ref{sec:Experiments} presents our experimental setup, execution, and results, highlighting the efficacy of our approach. Finally, Section~\ref{sec:Conclusion} concludes the paper with a summary of our findings and directions for future research.

\section{Background}\label{sec:Background}

\subsection{Vulnerability Artifacts}

The Common Vulnerabilities and Exposures (CVE) system, developed by the Mitre Corporation, offers a standardized approach to naming vulnerabilities in software and hardware. Each CVE identifier uniquely catalogs security flaws, enabling clear communication within the cybersecurity community. Despite its widespread adoption, CVE alone does not assess the severity of vulnerabilities. To address this, the Common Vulnerability Scoring System (CVSS) was introduced, providing a method to evaluate and score the impact of vulnerabilities. The National Vulnerability Database (NVD) integrates CVE identifiers with CVSS scores, enhancing the management and understanding of security vulnerabilities \cite{2007Common}.

Cyber Threat Intelligence (CTI) is crucial for organizations to anticipate, prevent, and respond to cyber threats. It involves the collection, analysis, and interpretation of information regarding cybersecurity threats \cite{2015Who}. CTI is categorized into tactical, operational, and strategic intelligence, each serving different purposes, from identifying imminent threats to informing long-term security strategies.

\subsection{MITRE ATT\&CK Framework}

The MITRE ATT\&CK framework, established in 2013, provides a comprehensive knowledge base that maps attacker tactics and techniques across various stages of an attack lifecycle. It categorizes these into three matrices: Enterprise, Mobile, and ICS, employing the TTPs (Tactics, Techniques, and Procedures) methodology. This framework aids researchers and security professionals in identifying, understanding, and mitigating cyber threats.

Integrating NLP and AI into cybersecurity, specifically in the context of the MITRE ATT\&CK framework, presents challenges due to the variability in vulnerability descriptions and the complexity of threat intelligence. Efforts to improve this integration include using advanced NLP models like BERT for accurate classification of TTPs within the ATT\&CK framework \cite{alves2022leveraging}. This approach has shown significant improvements in accuracy compared to traditional methods, highlighting the potential of AI in enhancing cybersecurity analytics.

The Center for Threat-Informed Defense proposes a methodology \cite{MITRE2021MappingATTCK} using the MITRE ATT\&CK framework to detail the impact of vulnerabilities identified by CVE identifiers. Mapping CVEs to ATT\&CK techniques, this method enhances defenders' understanding of potential adversary exploitation of vulnerabilities. We aim to investigate automating this mapping process to improve strategic decision-making in security measures.

\section{Methodology}\label{sec:Methodology}

\begin{figure*}[htbp]
\centering
\includegraphics[width=\textwidth]{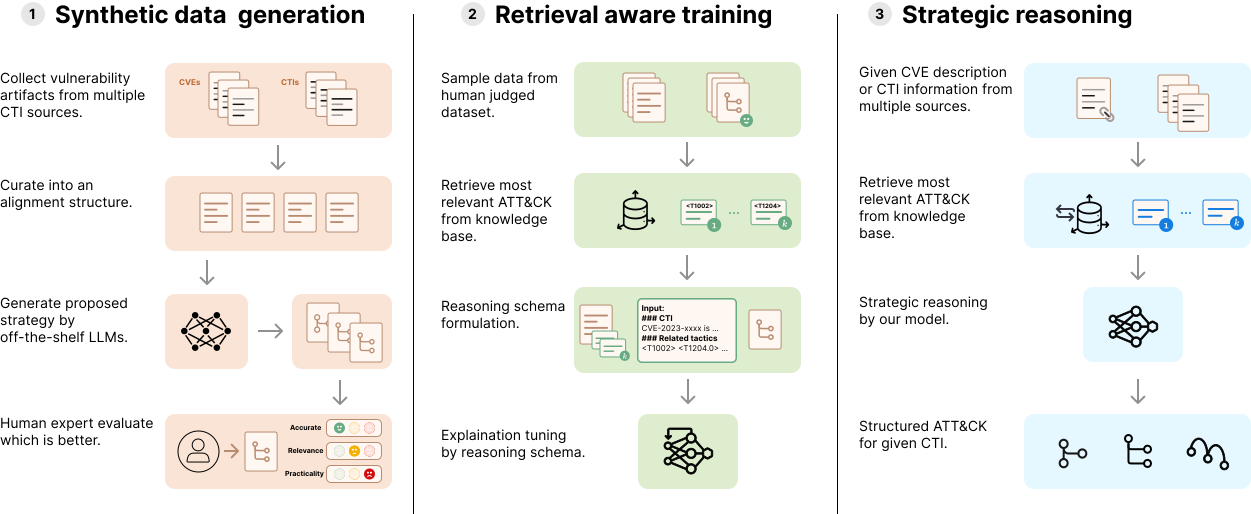}
\caption{A schematic overview of the three-phase methodology employed in our research: (1) Syntehtic data curation, which involves collating CVEs and CTIs from diverse sources, aligning them for LLM-generated strategic proposals, and with human expert feedback adjudication; (2) Retrieval-aware training, where we sample human-assessed data sets, retrieve relevant ATT\&CK information from a knowledge base, develop reasoning schemas, and refine explanation outputs; (3) Strategic reasoning, in which LLMs apply these schemas to deduce ATT\&CK techniques from CVEs and CTIs, aiding in the anticipation of cyber threats. Further details on the process are provided in Section 3 of our paper.}
\label{fig:methodology}
\end{figure*}

\subsection{Strategic Reasoning}

Strategic reasoning in cybersecurity involves the anticipatory analysis and decision-making processes to counteract vulnerabilities and cyber threats effectively. It is a critical component in understanding the complex interactions between different attack vectors and their impacts on cybersecurity infrastructure. Through strategic reasoning, defenders can prioritize response efforts, tailor defenses to specific threats, and enhance overall security posture. The following discussion illustrates the application of strategic reasoning in dissecting a significant cybersecurity vulnerability, leveraging the MITRE ATT\&CK framework for a structured analysis.

\begin{figure}[htbp]
\centering
\includegraphics[width=\columnwidth]{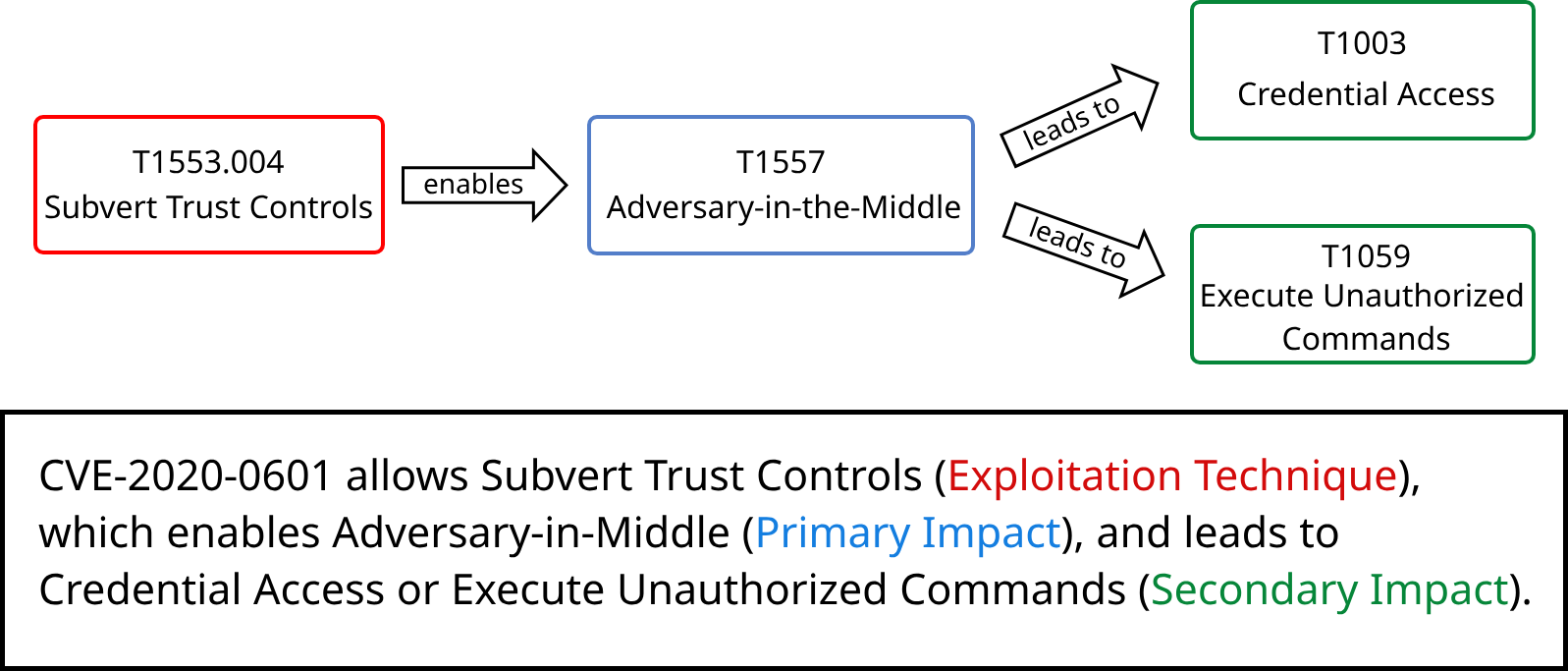}
\caption{CVE-2020-0601 (ChainOfFools), with it's related techniques and impacts.}
\label{fig:motivation}
\end{figure}

Consider CVE-2020-0601, a significant vulnerability in Windows CryptoAPI, also termed "ChainOfFools" or "CurveBall." This vulnerability compromises the cryptographic trust validation within Windows, allowing an attacker to impersonate trusted entities. Our approach employs the MITRE ATT\&CK framework to deconstruct this vulnerability, as visualized in Fig~\ref{fig:motivation}, and is detailed as follows:

\begin{itemize}
  \item \textbf{Exploitation Technique:} The attacker exploits 'Subvert Trust Controls' (T1553.004), targeting the vulnerability in cryptographic certificate validation.
  \item \textbf{Primary Impact:} Successful exploitation results in 'Man-in-the-middle' (T1557) attacks, where the attacker can intercept and manipulate communications.
  \item \textbf{Secondary Impact:} The breach of trust could lead to 'Credential Access' (T1003) or 'Execute Unauthorized Commands' (T1059), thereby further undermining the integrity of the system.
\end{itemize}

Mapping vulnerabilities to precise ATT\&CK tactics and techniques provides a clear and standardized framework to foresee and comprehend the impact of vulnerabilities. Our methodology's core aim is to enable high-level strategic planning, enhancing the ability to manage cybersecurity threats proactively. An schematic overview of our method showed in Fig~\ref{fig:methodology}.

\subsection{Dataset Collection}

Utilizing synthetic data for pretraining and instruction-tuning is pivotal in advancing cybersecurity strategic reasoning. We have developed a systematic approach to construct a synthetic cybersecurity dataset, tapping into the capabilities of Large Language Models (LLMs) such as Claude2 and GPT-4.

We obtained vulnerability information from \textit{cvelistV5}\footnote{\label{fn:cve-url}\url{https://github.com/CVEProject/cvelistV5}}, hosting over 200,000 CVE Records from 1999 to the present, forming the foundation of our synthetic dataset. For impact analysis, we utilized the MITRE ATT\&CK framework through \textit{mitreattack-python}\footnote{\label{fn:attck-url}\url{https://github.com/mitre-attack/mitreattack-python}}, primarily to validate LLM-generated records against the ATT\&CK framework's techniques or sub-techniques by ID and Name, avoiding hallucinations or kinks.

Advanced prompting strategies, particularly highlighted by the Medprompt study\cite{nori2023generalist}, have shown exceptional efficacy in vertical domains. We design domain-specific prompts for LLM (specifically GPT-4) to generate the insights that follows the MITRE ATT\&CK framework, enabling strategic CVE to ATT\&CK mapping within the framework. LLM input are formatted by the affected components and descriptions in CVE data, which is crucial as CTI and CVE data are often unstructured, require emphasis on descriptive elements to enhance the model's generalization ability. 

\begin{figure}[htbp]
\centering
\includegraphics[width=0.85\columnwidth]{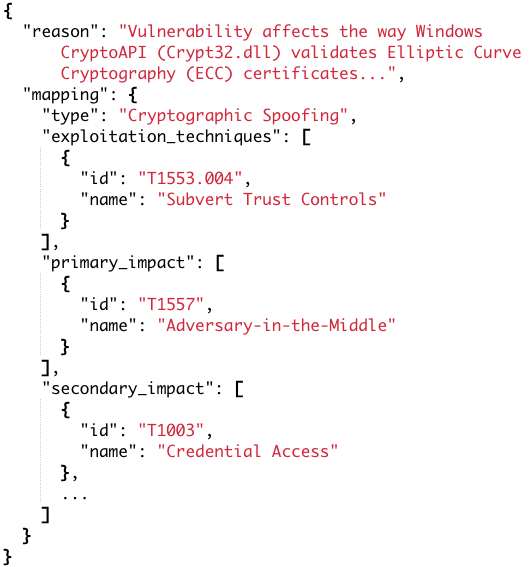}
\caption{CVE Mapping (CVEM) schema}
\label{fig:schema}
\end{figure}

The CVE-ATT\&CK Mapping Schema (CVEM), shown in Fig~\ref{fig:schema}, is introduced as output schema, enhances our model's strategic reasoning. It organizes techniques into a structured format, includes a "reason" attribute for interpretability, and employs reasoning prefixes for depth. These features collectively improve the model's output, making it more actionable for cybersecurity purposes. The impact of CVEM on model performance is detailed in Section~\ref{sec:Experiments}.

We involved a panel of cybersecurity experts to assess the generated CVEM for accuracy, relevance, and practicality. These mappings were rated as "Good," "Normal," or "Bad," ensuring the dataset's integrity and utility for improving cybersecurity measures. Mappings that achieved a "Good" rating across all aspects were classified as Human-curated CVEM, indicating their superior quality and pertinence to cybersecurity initiatives. Table~\ref{tab:cve-summary} summarizes these evaluations, with a specific focus on the Human-curated records. The 'Raw' column presents the initial number of CVEM generated; 'Accurate' denotes mappings validated for precision against the ATT\&CK framework, and 'Curated' reflects the subset of mappings deemed high-quality by expert review. 

\begin{table}[htbp]
\centering
\caption{Summary of CVEM dataset}
\label{tab:cve-summary}
\small 
\begin{tabular}{ccccc}
\toprule
\multicolumn{1}{c}{\multirow{2}{*}{Year}} & \multicolumn{1}{c}{\multirow{2}{*}{\#CVE}} & \multicolumn{3}{c}{\#CVE Mapping}  \\
\cmidrule(lr){3-5}
& & Raw & Accurate & Curated \\
\midrule
2015 & 8734 & 4058 & 1563 & 83  \\
2016 & 10538 & 4920 & 1998 & 80 \\
2017 & 16899 & 8052 & 3213 & 70 \\
2018 & 17254 & 10689 & 3823 & 61 \\
2019 & 16811 & 11399 & 4335 & 220 \\
2020 & 20400 & 13823 & 5441 & 210 \\
2021 & 21962 & 15257 & 5565 & 202 \\
2022 & 24280 & 16560 & 6139 & 202 \\
2023 & 18963 & 12869 & 4648 & 82 \\
\midrule
Total & 155841 & 97627 & 40806 & 1212 \\
\bottomrule
\end{tabular}
\end{table}

\subsection{Crimson}

Crimson targets elevating the strategic insight of LLMs within cybersecurity, leveraging fine-tuning techniques to bolster LLMs and embedding models. It strategically aims to integrate vulnerability management, threat modeling, and compensating controls, offering a comprehensive high-level strategic perspective.

\begin{figure}[htbp]
\centering
\includegraphics[width=0.85\columnwidth]{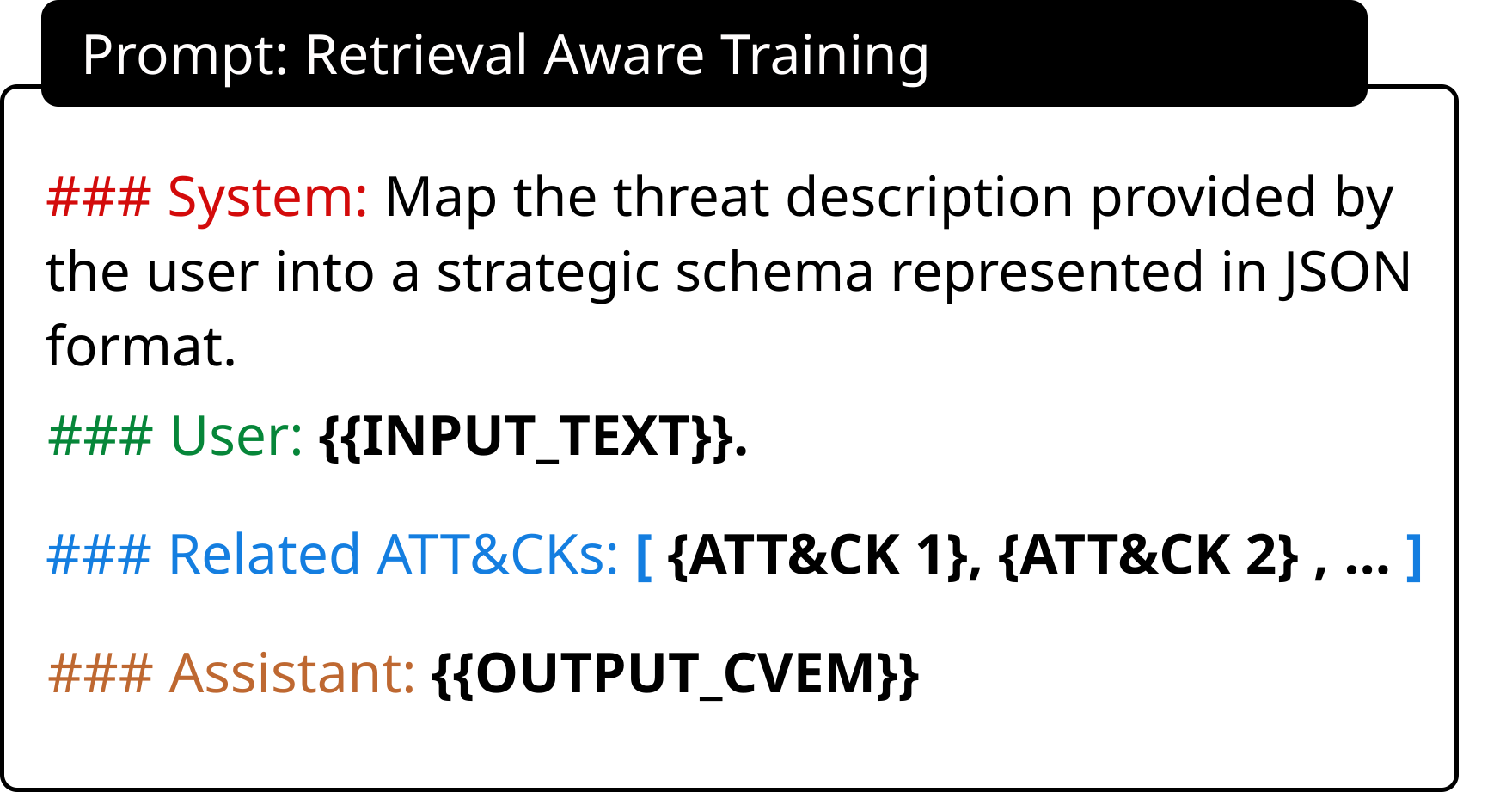}
\caption{Prompt template for RAT. Text in red indicates system-generated prompts, text in green represents user input, and text in blue signifies content retrieved by the system.}
\label{fig:prompt_template}
\end{figure}

\paragraph{Retrieval-Aware Training (RAT) and RAT-R.}
Central to Crimson's methodology is Retrieval Augmentated Generation (RAG)\cite{lewis2021retrievalaugmented}, which aims. For training a model which is retrieval aware on certain tasks, we introduced Retrieval-Aware Training (RAT) \cite{asai2022taskaware,patil2023gorilla}, RAT integrates real-time data retrieval into LLM training, enriching both training and inference phases with current and relevant cybersecurity data, as illustrated in Fig~\ref{fig:prompt_template}. 

Building on RAT, RAT-R introduces the "reason" property from the CVE-ATT\&CK Mapping Schema (CVEM) into the training process, enhancing the model's interpretability and strategic coherence.

Benefits of RAT and RAT-R include:
\begin{itemize}
    \item \textbf{Current and Contextual Information Integration:} Ensuring the model is up-to-date with the latest cybersecurity trends and data.
    \item \textbf{Enhanced Learning Performance:} Immersing the model in a context-rich environment improves its understanding and navigation of cyber threats.
    \item \textbf{Increased Output Reliability:} Utilizing current information reduces the likelihood of producing inaccurate or irrelevant outputs.
\end{itemize}

When combined with the CVE-ATT\&CK Mapping Schema (CVEM), RAT empowers LLMs to more accurately recognize and reflect the sequence of dependencies and operational constraints characteristic of cybersecurity strategies. This synergetic integration not only enhances strategic reasoning within the realm of cybersecurity but also establishes a more robust framework for the development of advanced defensive mechanisms.

\paragraph{Domain-specific Embedding Model.} Embedding models play a crucial role in distilling complex cyber threat data into a format that is both visually intuitive and strategically insightful for cybersecurity analysis. These models convert high-dimensional data into a simpler visual representation, clarifying semantic similarities and execution distinctions among attack techniques. For instance, while both T1189 (Drive-by Compromise) and T1190 (Exploit Public-Facing Application) methods enable initial access by exploiting vulnerabilities, their approaches diverge: T1189 adopts an opportunistic strategy dependent on user interaction, whereas T1190 employs a targeted attack on specific public-facing applications. Fine-tuning an embedding model with approximately 40,000 CVEM records significantly enhances the relevance of these distinctions, aiding experts in distinguishing between seemingly similar attack strategies with greater precision.

\begin{figure}[htbp]
\centering
\includegraphics[width=0.85\columnwidth]{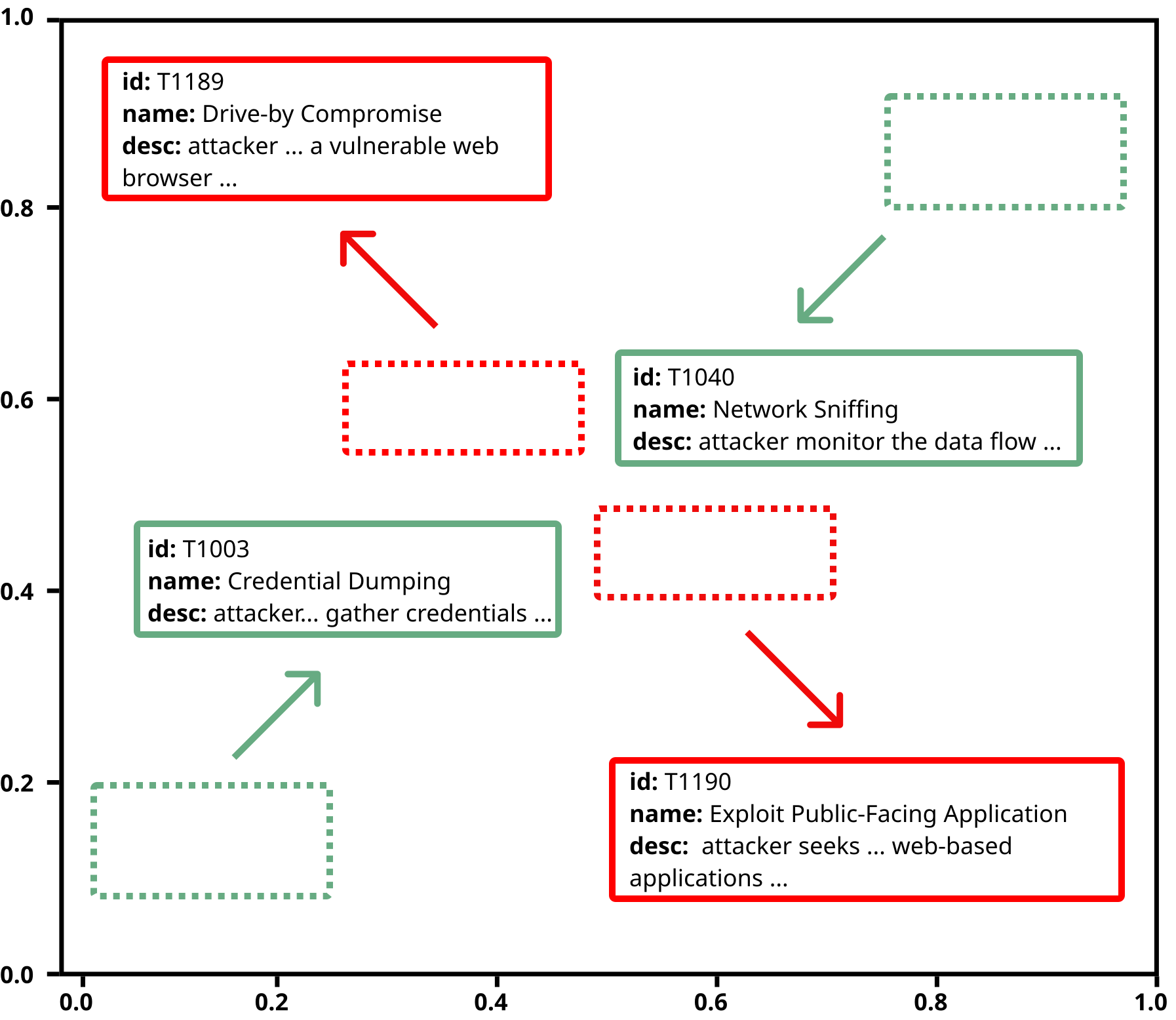}
\caption{t-SNE Visualization of Cybersecurity Techniques with Contextual Differentiation. This figure demonstrates the embeddings for two technique pairs: T1189 and T1190, highlighted with red rectangles, alongside T1003 and T1040 as green rectangles. Rectangles without borders depict pre-contextual differentiation embeddings, while those with solid borders illustrate post-domain-specific modeling adjustments.}
\label{fig:embedding}
\end{figure}

In Fig~\ref{fig:embedding}, we showcase the embeddings of two ATT\&CK technique pairs: T1189 and T1190, and T1003 and T1040. Initially, the base embedding model places T1189 and T1190 close together in the embedding space, grouping them by their basic semantic meanings. However, after fine-tuning with cybersecurity-specific contextual data (CVEM), the distance between T1189 and T1190 increases, reflecting their distinct cybersecurity implications as discussed earlier. Similarly, T1003 (Credential Dumping) and T1040 (Network Sniffing), despite differing semantically, both relate to user credential collection activities. Thus, in the fine-tuned embedding space, they are brought closer, aligning them more closely in their decision-making context.

Fine-tuning embedding models with targeted cybersecurity data markedly improves the ability to identify and address the complex subtleties of cyber threats. A quantitative analysis of these experiments will be detailed in Section~\ref{sec:Experiments}.

\subsection{Strategy Measurement}

In our research, we focus on metrics tailored to the cybersecurity domain, essential for assessing a model's precision in mapping CVEs to ATT\&CK techniques and its effectiveness in interpreting complex cybersecurity situations. Our evaluation framework simulates real-world cybersecurity challenges, crucial for verifying the model's resilience and practical applicability.

\paragraph{F1-Score and IOU.} Despite Crimson being a language model that produces a finite set of ATT\&CK techniques in a tree structure, standard classification metrics, like F1-Score and Intersection over Union (IOU), remain applicable to our task. The F1-Score balances precision in identifying accurate threats with recall's emphasis on coverage, reflecting Crimson's strategic discernment in threat prioritization. The IOU quantifies the alignment between Crimson's predictions and validated threat mappings, showcasing its strategic accuracy in navigating the cybersecurity landscape. These metrics are integral for appraising Crimson's strategic reasoning, crucial for effective threat identification and mitigation.

\paragraph{Hallucination and Error Metrics.} Addressing the unique challenges posed by LLMs, we introduce metrics for hallucination and error rates. A \textit{hallucination} occurs when Crimson predicts primary or secondary impacts as ATT\&CK techniques that do not exist—effectively creating fictitious techniques. In contrast, an \textit{error} refers to incorrect CVEM format, diverging from valid ATT\&CK techniques organizing format. 

\begin{figure}[htbp]
\centering
\includegraphics[width=\columnwidth]{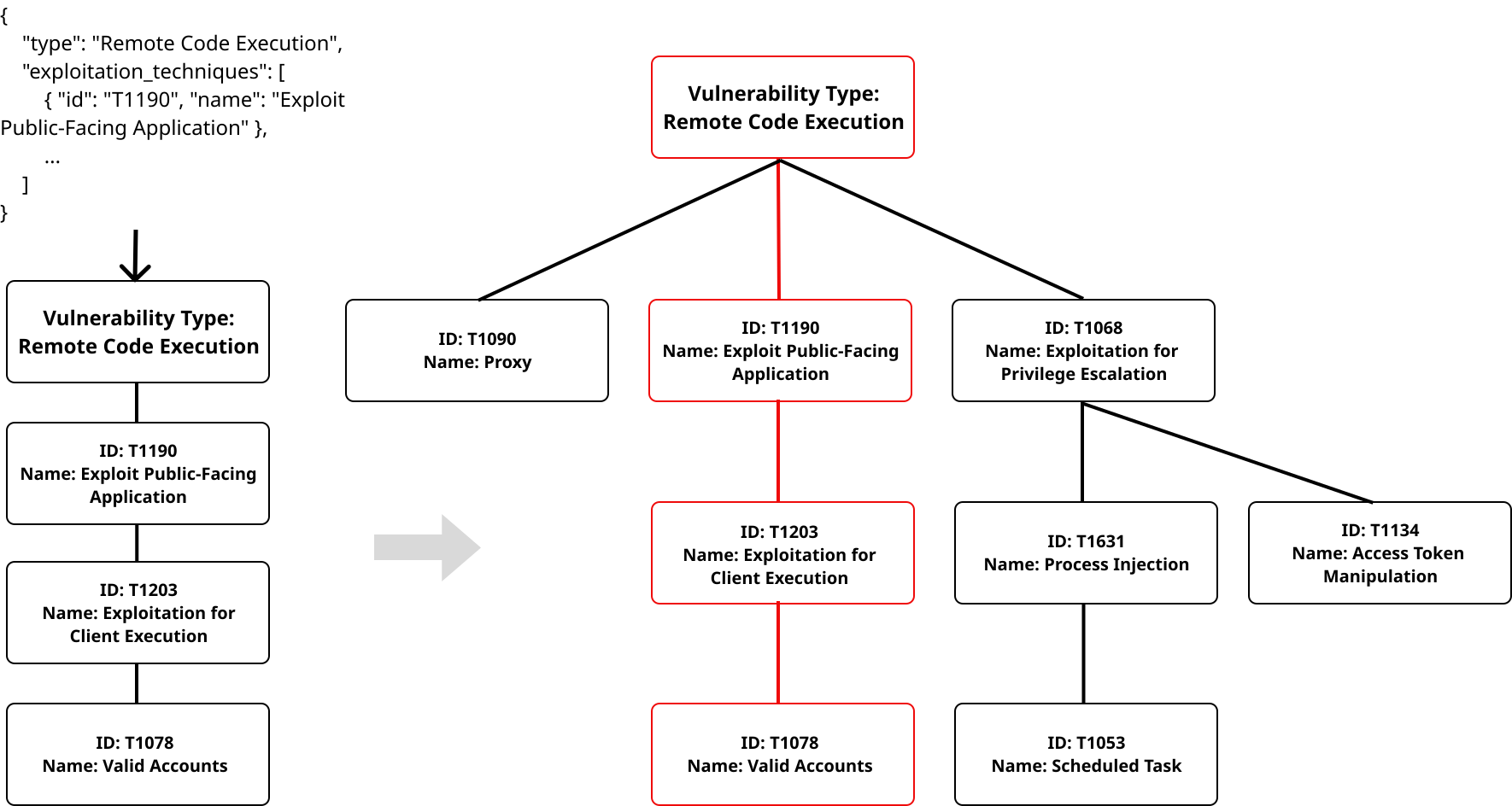}
\caption{AST Sub-Tree Matching for ATT\&CK Technique Synthesis. }
\label{fig:ast_subtree}
\end{figure}

\paragraph{AST Sub-Tree Matching.} We introduced Abstract Syntax Trees (ASTs) in our analysis for their nuanced examination of source code's syntactic structure. By adopting ASTs for our CVE-ATT\&CK Mapping Schema (CVEM), we leverage their capability to outline the intricate relationships between different cybersecurity actions and impacts. This method allows for a detailed evaluation of strategic decisions, surpassing the general overview provided by F1-Score and IOU. The AST accuracy, illustrated in Figure~\ref{fig:ast_subtree}, is a fine-grained metric assessing the model's step-by-step strategic alignment with ATT\&CK techniques, offering a deeper insight into its semantic precision and the logical coherence of its threat assessments. This detailed approach underscores the model's proficiency in executing structured cybersecurity strategies, a critical aspect for practical defense mechanisms. 

As illustrated in Figure~\ref{fig:ast_subtree}, we validate the model's output through visual comparison with a predefined hierarchy of ATT\&CK techniques. This analysis examines the model's strategic accuracy by aligning it with the systematic logic inherent in cyber-attack strategies. By matching each node to specific ATT\&CK techniques, identified by unique IDs and terms, we construct a layered view of potential exploitation paths. The model's performance is evaluated based on its ability to accurately reconstruct the AST, thereby using AST accuracy as a metric for semantic precision in cybersecurity threat analysis. The calculation of AST accuracy detailed in Algorithm~\ref{alg:ast_coverage_accuracy}. 

\begin{algorithm}
\caption{Calculate AST Accuracy}
\label{alg:ast_coverage_accuracy}
\begin{algorithmic}[1]
\REQUIRE Model output $M$, Ground truth $G$
\ENSURE AST accuracy $A$
\STATE Let $K = \text{keys}(M) \cup \text{keys}(G)$
\STATE Initialize $total\_score = 0$, $max\_score = 0$
\FORALL{$k \in K$}
    \STATE Let $I = \text{ids}(M[k]) \cap \text{ids}(G[k])$
    \STATE Let $U = \text{ids}(M[k]) \cup \text{ids}(G[k])$
    \STATE $S_k \gets (|U| > 0) ? \frac{|I|}{|U|} : 0$
    \STATE $max\_score \gets max\_score + 1$
    \STATE $total\_score \gets total\_score + S_k$
\ENDFOR
\STATE $A \gets (max\_score > 0) ? \frac{total\_score}{max\_score} : 0$
\RETURN $A$
\end{algorithmic}
\end{algorithm}

\begin{figure*}[htbp]
\centering
\includegraphics[width=\textwidth]{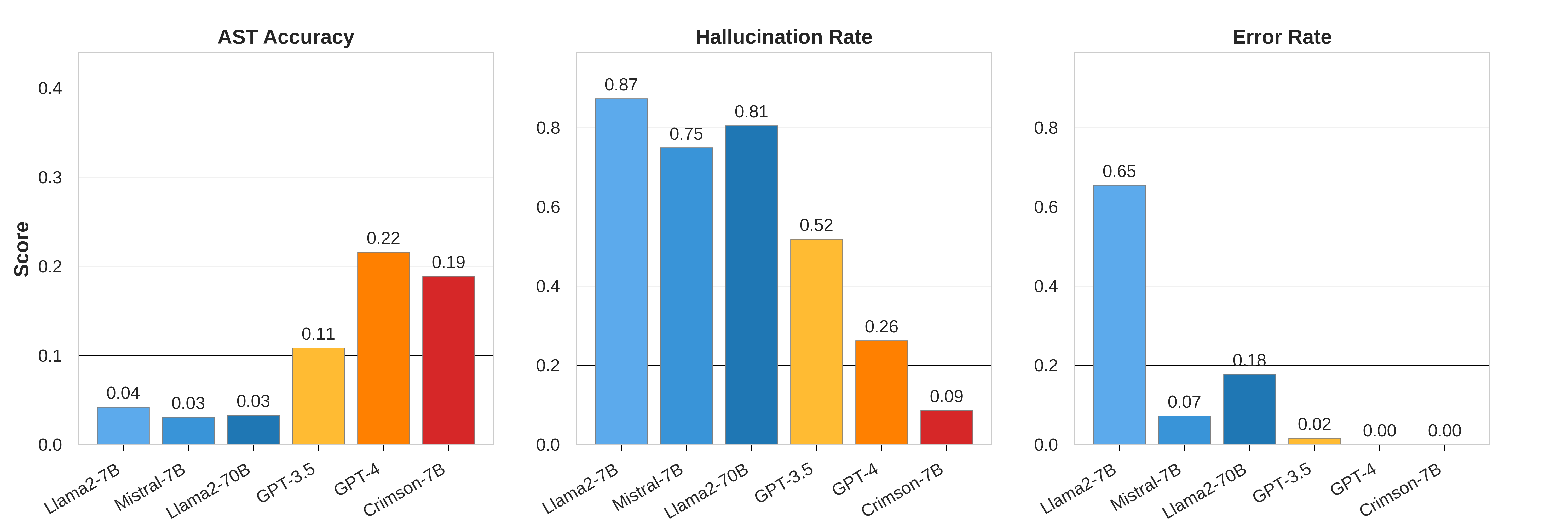}
\caption{Performance Metrics of Language Models in CVE-to-ATT\&CK Mapping. The chart evaluates AST Accuracy, Hallucination Rate, and Error Rate across models like Llama2-7B, Mistral-Instruct-7B, Llama2-70B, GPT-3.5, GPT-4, and Crimson-7B. The metrics assess each model's ability to accurately parse CVEs and predict ATT\&CK techniques, with lower hallucination and error rates indicating better performance.}
\label{fig:model_performance}
\end{figure*}

\begin{table*}[h]
    \centering
    \caption{Impact on Model Performance Metrics (Model Scale, Retriever, \& Prompt Method)}
    \label{tab:model_metrics}
    \begin{adjustbox}{max width=\textwidth} 
        \begin{tabular}{llllcccccc}
            \toprule
            \textbf{Model} & \textbf{Retriever} & \textbf{Prompts} & \multirow{2}{*}{\makecell[l]{\textbf{AST} \\ \textbf{Accuracy}}} & \multicolumn{3}{c}{\textbf{F1 Score}} & \multicolumn{3}{c}{\textbf{IoU Score}} \\
            \cmidrule(lr){5-7} \cmidrule(lr){8-10}
             &  &  &  & \textbf{ET} & \textbf{PI} & \textbf{SI} & \textbf{ET} & \textbf{PI} & \textbf{SI} \\
            \midrule
            Llama-2-7B & - & Few-shot & 0.0421 & 0.0 & 0.0 & 0.0 & 0.0825 & 0.019 & 0.0194 \\
            Mistral-7B-Instruct & - & Few-shot & 0.0309 & 0.0 & 0.0323 & 0.0 & 0.0094 & 0.0403 & 0.0188 \\
            Llama2-70B & - & Few-shot & 0.033 & 0.0 & 0.032 & 0.0027 & 0.032 & 0.08 & 0.0121 \\
            gpt-3.5-turbo & - & Few-shot & 0.1085 & 0.0165 & 0.0718 & 0.0159 & 0.0427 & 0.1063 & 0.0221 \\
            gpt-4-1106-preview & - & Few-shot & \textbf{0.2157} & \textbf{0.2579} & \textbf{0.122} & \textbf{0.0716} & \textbf{0.2745} & \textbf{0.1645} & \textbf{0.0706} \\
            \midrule
            Crimson-7B & BGE@10 & RAT-R & 0.1425 & 0.0285 & 0.0033 & 0.0 & 0.0307 & 0.0033 & 0.0 \\
            Crimson-7B & Crimson@10 & RAT & 0.1826 & 0.2155 & 0.023 & \textbf{0.0329} & 0.1582 & 0.0263 & \textbf{0.0362} \\
            Crimson-7B & Crimson@10 & RAT-R & \textbf{0.1888} & \textbf{0.2264} & \textbf{0.0329} & 0.0296 & \textbf{0.1697} & \textbf{0.0362} & 0.0329 \\
            \midrule
            Crimson-70B & BGE@10 & RAT-R & 0.154 & 0.0389 & 0.0033 & 0.0 & 0.037 & 0.0033 & 0.0 \\
            Crimson-70B & Crimson@10 & RAT & \textbf{0.2094} & \textbf{0.2708} & \textbf{0.0296} & \textbf{0.0362} & \textbf{0.1996} & \textbf{0.0362} & \textbf{0.0395} \\
            Crimson-70B & Crimson@10 & RAT-R & 0.2046 & 0.2471 & 0.0263 & 0.0329 & 0.1804 & 0.0296 & 0.0362 \\
            \midrule
            Crimson-7B & Crimson@100 & RAT-R & 0.6841 & 0.7203 & 0.5132 & 0.6139 & 0.7991 & 0.6579 & 0.7063 \\ 
            Crimson-70B & Crimson@100 & RAT-R & \textbf{0.6992} & \textbf{0.7291} & \textbf{0.5362} & \textbf{0.6316} & \textbf{0.8196} & \textbf{0.6908} & \textbf{0.7467} \\
            \bottomrule
        \end{tabular}
    \end{adjustbox}
\end{table*}

\begin{table*}[h]
    \centering
    \caption{Impact on Embedding Model Performance Metrics}
    \label{tab:embedding_model_metrics}
    \begin{adjustbox}{max width=\textwidth}
        \begin{tabular}{lcccccccc}
            \toprule
            \textbf{Model Name} & \textbf{MRR@10} & \textbf{MAP@100} & \multicolumn{2}{c}{\textbf{Accuracy}} & \multicolumn{2}{c}{\textbf{Precision}} & \multicolumn{2}{c}{\textbf{Recall}} \\
            \cmidrule(lr){4-5} \cmidrule(lr){6-7} \cmidrule(lr){8-9}
            & & & @1 & @5 & @1 & @5 & @1 & @5 \\
            \midrule
            bge-large-en-v1.5 & 0.167 & 0.077 & 0.077 & 0.281 & 0.077 & 0.067 & 0.016 & 0.066 \\
            *crimson-embedding & 0.413 & 0.139 & 0.302 & 0.561 & 0.302 & 0.142 & 0.059 & 0.141 \\
            \bottomrule
        \end{tabular}
    \end{adjustbox}
\end{table*}

\section{Experiments}\label{sec:Experiments}

\subsection{Evaluation Setup}

\paragraph{Test Bed.} 
Our experiments are conducted on a GPU server facilitated by RunPod, equipped with 4 NVIDIA H100 GPUs (80G PCIE), providing a total of 320 GB VRAM and supported by 704 GB of system RAM.

\paragraph{Datasets.} 
We utilize two datasets for our experiments: CVEM-40k, consisting of 40,806 accurately curated records, and CVEM-1k, comprising 1,212 human-curated records, as detailed in Table~\ref{tab:cve-summary}. The CVEM-1k dataset, split 80-20 for training and validation, is formatted following OpenAI's chat-based fine-tuning protocol, delineating roles into "system," "user," and "assistant" to facilitate instructions, CVE/CTI descriptions, and CVEM ground truths, respectively. 

\paragraph{Models.} 
In our research, we utilize advanced open-source language models: Mistral\cite{jiang2023mistral}, LLaMA-2\cite{touvron2023llama}, and bge-large-en-v1.5\cite{bge_embedding} as our retriever. Specifically, we fine-tune LLaMA-2 models employing the Low-Rank Adaptation (LoRA)\cite{hu2021lora} technique, configured with a rank of 8, alpha of 16, across all dense layers, and a base learning rate of $1 \times 10^{-4}$. This fine-tuning leverages the TorchTrainer framework alongside DeepSpeed ZeRO-3 optimization, focusing on the 7B and 70B versions of LLaMA-2. Additionally, the bge-large-en-v1.5 model undergoes fine-tuning on the CVEM-40k dataset, enhancing its retrieval capabilities for cybersecurity applications. We further explore the impact of varying the number of top-N retrieved results during context setting, specifically @10 and @100, to assess performance changes.

Model configurations and performance benchmarks are provided in Table~\ref{tab:model_metrics}.

\subsection{Overall Performance}
Our evaluation, as visualized in Figure~\ref{fig:model_performance}, demonstrates that fine-tuning methodologies and model scale significantly impact AST accuracy and the ability to parse CVEs effectively.

The GPT-4 model exhibits superior performance in few-shot settings, outperforming all open-source LLMs and highlighting its robustness in strategic reasoning tasks.

Open-source LLMs, in the absence of Retrieval-Aware Training (RAT), show poor AST accuracies. While scale does not necessarily correlate with accuracy, it contributes to instruction adherence, also reduced hallucination and error rates.

The deployment of Retrieval-Augmented Training (RAT) and its enhanced iteration, RAT-R, markedly boosts the capabilities of smaller language models, elevating them to rival their larger counterparts in performance. This advancement is crucial for generating dependable CVE-to-ATT\&CK mappings, smaller model also offering advantages in both cost-efficiency and processing throughput.

\subsection{Ablation Study}

The ablation study, as detailed in Table~\ref{tab:model_metrics}, meticulously examines the influence of model scale, retriever integration, and the application of Retrieval-Aware Training (RAT) and its advanced iteration, RAT-R, on a spectrum of performance metrics. This analysis is pivotal for understanding the nuanced contributions of each component to the overall effectiveness of our strategic reasoning framework within cybersecurity contexts.

\paragraph{Model Scale and Performance.} Our findings underscore the substantial impact of model scale on strategic reasoning capabilities. Notably, the transition from smaller-scale models like Llama-2-7B to larger counterparts such as GPT-4 significantly enhances AST accuracy and F1 scores. However, the integration of RAT-R with the Crimson model, even at a smaller scale (7B), bridges this performance gap, closely approaching or even matching the strategic reasoning prowess of GPT-4. This observation highlights the potential of advanced training methodologies to amplify the strategic reasoning capabilities of smaller models, making them viable contenders against larger models in strategic reasoning tasks.

\paragraph{Enhancing Model Interpretability and Contextuality.} The introduction of a "reason" attribute within the output schema through the RAT-R process marks a significant advancement in model interpretability and the contextuality of outputs. This enhancement not only facilitates a deeper understanding of the model's reasoning process but also contributes to a marked improvement in AST accuracy and a reduction in hallucination rates. This advancement illustrates the critical role of output schema design in enhancing the strategic reasoning performance of language models.

\paragraph{Retriever Integration and Retrieval Scope.} Employing the domain-specifically fine-tuned *crimson-embedding model has shown a pronounced improvement in retrieval related metrics, including MRR and MAP, and also enhances the model's reasoning ability during the inference process. These adjustments underscore the importance of precision in information retrieval and the integration of contextually relevant data during the reasoning process.

Expanding retrieval windows, such as adjusting the retrieval scope to include the top 100 related ATT\&CK techniques (@100), further accentuates the benefits of domain-specific fine-tuning. This approach suggests that optimizing the embedding model could indirectly improve inference performance, presenting a more cost-effective alternative compared to extensive fine-tuning of language models.

\paragraph{Strategic Implications.} The findings from our ablation study, particularly the efficacy of RAT and RAT-R in enhancing the strategic reasoning capabilities of language models, suggest a paradigm shift in how we approach model training and deployment in cybersecurity contexts. By focusing on the integration of advanced training methodologies, domain-specific fine-tuning, and precision in information retrieval, we can significantly enhance the strategic reasoning capabilities of models at a fraction of the computational cost and complexity associated with scaling up model size.

In conclusion, our ablation study not only sheds light on the individual and collective impacts of model scale, retriever integration, and advanced training methodologies on model performance but also sets the stage for future explorations into optimizing language models for strategic reasoning in cybersecurity and beyond.

\section{Conclusion}\label{sec:Conclusion}

Our research confirms the feasibility of using Large Language Models to provide actionable mappings from CVEs to ATT\&CK techniques, thus offering a strategic advantage in cybersecurity defense. The fine-tuning approaches, Retrieval-Aware Training (RAT) and its refined version, RAT-R, are shown to significantly enhance the reasoning capabilities of LLMs, ensuring the reliability of threat predictions. 

The experimental results underline the importance of domain-specific fine-tuning, retrieval enhancement, and the integration of a reasoning schema for improving the accuracy and interpretability of model predictions. Future work should focus on refining retrieval mechanisms and exploring the scalability of these models to maintain efficacy as the cyber threat landscape evolves.

{\footnotesize \bibliographystyle{unsrt}
\bibliography{references}}

\end{document}